\documentclass{ws-procs10x7}
\usepackage{psfig}

\newcolumntype{d}[1]{D{.}{.}{#1}}

\def\Journal#1#2#3#4{{\it #1} {\bf #2}, #3 (#4)}
\def\etal{{\sl et al.}} 
\newcommand{\be}{\begin{equation}}
\newcommand{\ee}{\end{equation}}
\newcommand{\bea}{\begin{eqnarray}}
\newcommand{\eea}{\end{eqnarray}}
\newcommand{\xf}{x_{\mathrm F}}
\newcommand{\PAN}{Phys.\ Atom.\ Nucl.}
\newcommand{\YAF}{Yad.\ Phys.}
\newcommand{\rar}{\rightarrow}
\newcommand{\pdup}{p_\uparrow}
\newcommand{\pimp}{\pi^- + \pdup \rar \pi^0 + X}
\newcommand{\pdupp}{\pdup + p \rar \pi^0 + X}
\newcommand{\ppdup}{p+\pdup \rar \pi^0 + X}
\newcommand{\PLB}{Phys.\ Lett.}
\newcommand{\PRD}{Phys.\ Rev.}
\makeindex
\begin{document}

\title{Single spin asymmetries in inclusive $\pi^0$ production 
in $p+p$ and $\pi^-+p$ interactions at 40-70 GeV}

\author{V.V.~MOCHALOV$^*$, A.M.~DAVIDENKO, A.A.~DEREVSCHIKOV, 
Y.M.~Goncharenko, V.Y.~KHODYREV, V.I.~KRAVTSOV, 
Y.A.~MATULENKO, Y.M.~MELNICK, A.P.~MESCHANIN, N.G.~MINAEV, 
D.A.~MOROZOV, L.V.~NOGACH, S.B.~NURUSHEV, L.F.Prudkoglyad,
P.A.~SEMENOV, L.F.~SOLOVIEV, A.N.~VASILIEV 
\and A.E.~YAKUTIN}

\address{Institute for High Energy Physics, Protvino, 
Moscow region, 142281, Russia\\
$^*$E-mail: mochalov@ihep.su}

\author{N.L.~BAZHANOV, N.S.~BORISOV, A.N.~FEDOROV, 
V.G.~Koloimiets, A.B.~LAZAREV, 
A.B.~NEGANOV, Y.A.~PLIS, O.N.~SHCHEVELEV \and Y.A.~USOV}

\address{JINR, Dubna, Moscow region, Russia}

\twocolumn[\maketitle\abstract {We present recent results 
of single-spin asymmetry $A_N$ measurements in $\pi^0$ inclusive 
production. Asymmetry was measured in $\pi^-p$ and 
$pp$ interactions at 40 and 70 GeV correspondingly. 
Significant asymmetry was observed in the polarized target
fragmentation region. The results are in agreement with 
"universal threshold" of single-spin asymmetry.}]

\section{Introduction} 

\begin{figure*}
\centerline{\psfig{file=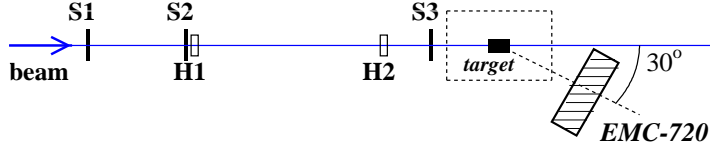,width=0.8\textwidth}}
\caption{Experimental Setup PROZA-M. S1-S3 -- trigger scintillation 
counters; H1-H2 -- hodoscopes; $EMC-720$ -- electromagnetic 
calorimeter; $target$ -- polarized target.}
\label{fig:setup}
\end{figure*}

Polarization experiments give us an unique opportunity to probe the 
nucleon internal structure. While spin averaged cross-sections can be 
calculated within acceptable accuracy, current theory of strong 
interactions can not describe single-spin asymmetries and 
polarization. Unexpected large values of single spin asymmetry 
in inclusive $\pi$-meson production are real challenge 
to current theory because naive perturbative Quantum Chromodynamics 
predicts small asymmetries decreasing with transverse momentum.

The asymmetry measurements in inclusive $\pi^0$-meson 
production with the use of pion and proton beams at 
the polarized target fragmentation region were carried out at 
the Protvino U-70 accelerator in the energy range between 40 and 
70 GeV to investigate the dependence of $A_N$ on a particle
flavor.

\section{Experimental Measurements}

\subsection{Experimental Setup}

The experiment was carried out at the PROZA-M experimental set-up 
(See {\bf Fig.~\ref{fig:setup}}).

\begin{figure}[b]
\vspace*{-6mm}
\centerline{\psfig{file=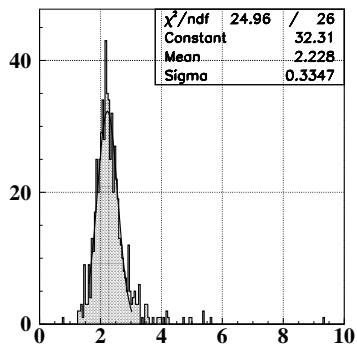,width=2.2in}}
\vspace*{-6mm}
\caption{Sensitivity of the EMC cells (in MeV/ADC count).}
\label{fig:sensitivity}
\end{figure}

A 70~GeV proton beam extracted from the U-70 main ring with the 
use of a bent crystal had intensity  $(3-6) \cdot 10^6$ 
protons/2 sec. spill. The full cycle time was 10 sec. 
A frozen polarized propane-diol ($C_3H_8O_2$) 
target was being operated with an average polarization of 85\%. 
Three scintillation counters $S1$--$S3$ and two two-coordinate 
hodoscopes $H1$--$H2$ were used for a beam particle registration 
and a zero level trigger. A first level analog trigger provided the 
events selection with the energy deposited into the 
calorimeter above 2 GeV. Gamma-quanta were detected by the 
electromagnetic calorimeter EMC (720 lead glass blocks) placed 
2.3 m downstream the target at $30^{\circ}$ respectively to the beam direction. 
The EMC was adjusted to measure low energy $\gamma$-quanta, 
starting from 100~MeV. The sensitivity of each cell 
was about 2.3~MeV/ADC count (See {\bf Fig.~\ref{fig:sensitivity}}), the width of the 
distribution was less than 15\%. The detailed description of the 
experimental setup is presented elsewhere\cite{setup}.

\subsection{Asymmetry calculation}

The single spin asymmetry $A_N$ is defined by the
expression given in the eq. (\ref{eq:asymdef}).

\bea
A_N(x_F,p_T)= \frac{1}{P_{targ}}\cdot 
\frac{1}{<cos\phi>}\times \nonumber \\
\times \frac{\sigma_{\uparrow}^H(x_F,p_T)-\sigma^H_{\downarrow}(x_F,p_T)}
{\sigma^H_{\uparrow}(x_F,p_T)+\sigma^H_{\downarrow}(x_F,p_T)}
\label{eq:asymdef}
\eea

\noindent
where $P_{targ}$ is the target polarization,  $\phi$ is the 
azimuthal angle between the target-polarization vector
and the normal to the plane spanned by the beam axis and the momentum 
of the outgoing neutral pion, and $d\sigma^H_{\uparrow}$ 
($d\sigma^H_{\downarrow}$) are the invariant differential cross
sections for a neutral-pion production on hydrogen for
the opposite directions of the target-polarization vector.
We detected neutral pions in the azimuthal 
angle range of $(180 \pm 15)^{\circ}$; therefore, we set $\cos \phi= -1$.  
Since the $\pi^0$'s detection efficiency is identical for the two 
directions of the target polarization vector, we find for 
the detector on the right side of the beam, that

\begin{figure}[t]
\centerline{\psfig{file=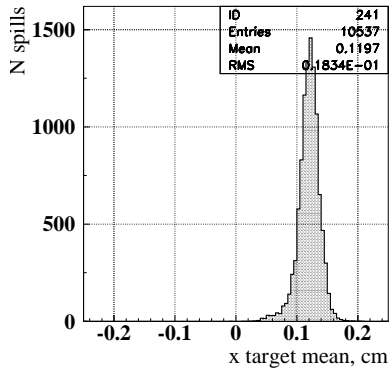,width=2.2in}}
\vspace*{-4mm}
\caption{Average beam position during spill.}
\label{fig:beam}
\end{figure}

{\small 
\be
A_N=-\frac{D}{P_{targ}}\cdot A_N^{raw} =
-\frac{D}{P_{target}}\cdot
\frac{n_{\uparrow}-n_{\downarrow}}{n_{\uparrow}+n_{\downarrow}}
\label{eq:asymreal}
\ee 
}

\noindent
where $A_N^{raw}$ is the raw asymmetry actually measured
in the experiment, $D$ is the target-dilution factor, and
$n_{\uparrow}$ ($n_{\downarrow}$) are the normalized 
(to the monitor) numbers of detected neutral pions for the 
up and down directions of the target-polarization vector. The procedure used
to calculate $D \approx  8.1 $ was described in detail 
elsewhere\cite{protv_yaf}. In measuring the asymmetry 
$A_N$, there can arise an additional asymmetry caused by 
a trigger-electronics jitter, failures of the monitor 
counters, a beam drift or by some other reasons. This gives 
a rise to a systematic bias of the true asymmetry. 
A method that can be used to remove this bias and 
which is based on the fact that the asymmetry of 
photon pairs off the neutral pion mass peak is zero 
is described in detail in\cite{protv40}. We estimated 
the stability of the EMC energy scale at the level 
of 0.1\% based on the stability of $\pi^0$ -mass. 
The beam position instability was the main reason of 
systematic bias of the true asymmetry in the previous 
measurements. The improvement of the beam extraction 
technique\cite{beam-aseev} allowed us to achieve the 
stability of a beam position at the target at the level 
of 0.2 mm (See {\bf Fig.~\ref{fig:beam}}). 

\section{Data Analysis and Results}

\subsection{$\pi^0$-reconstruction}

\begin{figure}
\centerline{\psfig{file=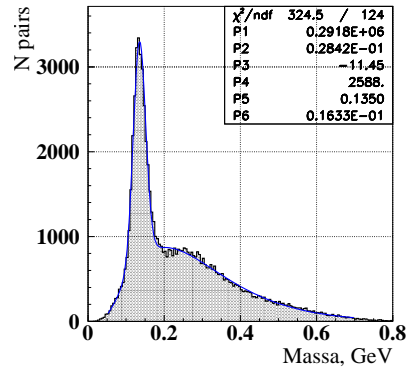,width=2.2in}}
\vspace*{-4mm}
\caption{Mass of the $\gamma \gamma$-pair for $-0.30 < \xf <-0.25$
interval.}
\label{fig:mass}
\end{figure}

The events with more than two clusters were selected 
for a $\pi^0$-search and analysis. The clusters were considered 
to be good if their shower shape is comparable with the 
shape of the electromagnetic shower. A special procedure has 
being developed for compensate energy losses at 
low energies\cite{protv40} and due 
to the shower leakage at large angles\cite{angle_sol}. 
The mass spectrum of $\gamma \gamma$-combination after 
all corrections is presented in {\bf Fig.~\ref{fig:mass}}.  
The mass width $\sigma$ is of the 
order of 16 MeV/c$^2$. The resolution is pretty good, in spite 
of the size of the target (about 20 cm long) is noticeable in 
comparison with the distance from the detector to the target 
(about 2.3 m). 

\begin{figure}[b]
\vspace*{-6mm}
\centerline{\psfig{file=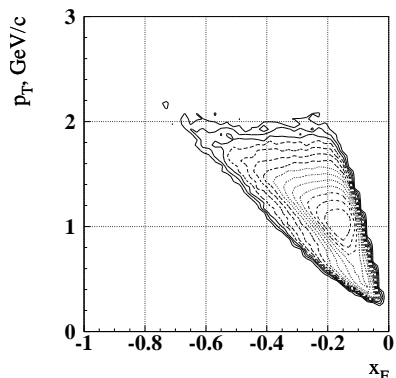,width=2.2in}}
\vspace*{-4mm}
\caption{Kinematic parameters of $\gamma \gamma$-pair for
mass region $110<m_{\gamma \gamma} (MeV/c^2)<160$.}
\label{fig:pt_xf}
\end{figure}

We have the possibility to measure $A_N$ in a wide 
interval of $-0.2>\xf>-0.8$. The kinematic parameters of 
the detected particles (transverse momentum $p_t$ and $\xf$)
are correlated. The two-dimensional distribution of the 
kinematic parameters of $\gamma \gamma$-combinations in the mass 
range $105<m_{\gamma \gamma} (MeV/c^2)<165$ for the 
recent data taking run are presented on 
{\bf Fig.~\ref{fig:pt_xf}}. We plan to 
increase statistics for large negative values of
$\xf$ over the next exposition.

\subsection{Measured Asymmetry}.

\begin{figure}[t]
\centerline{\psfig{file=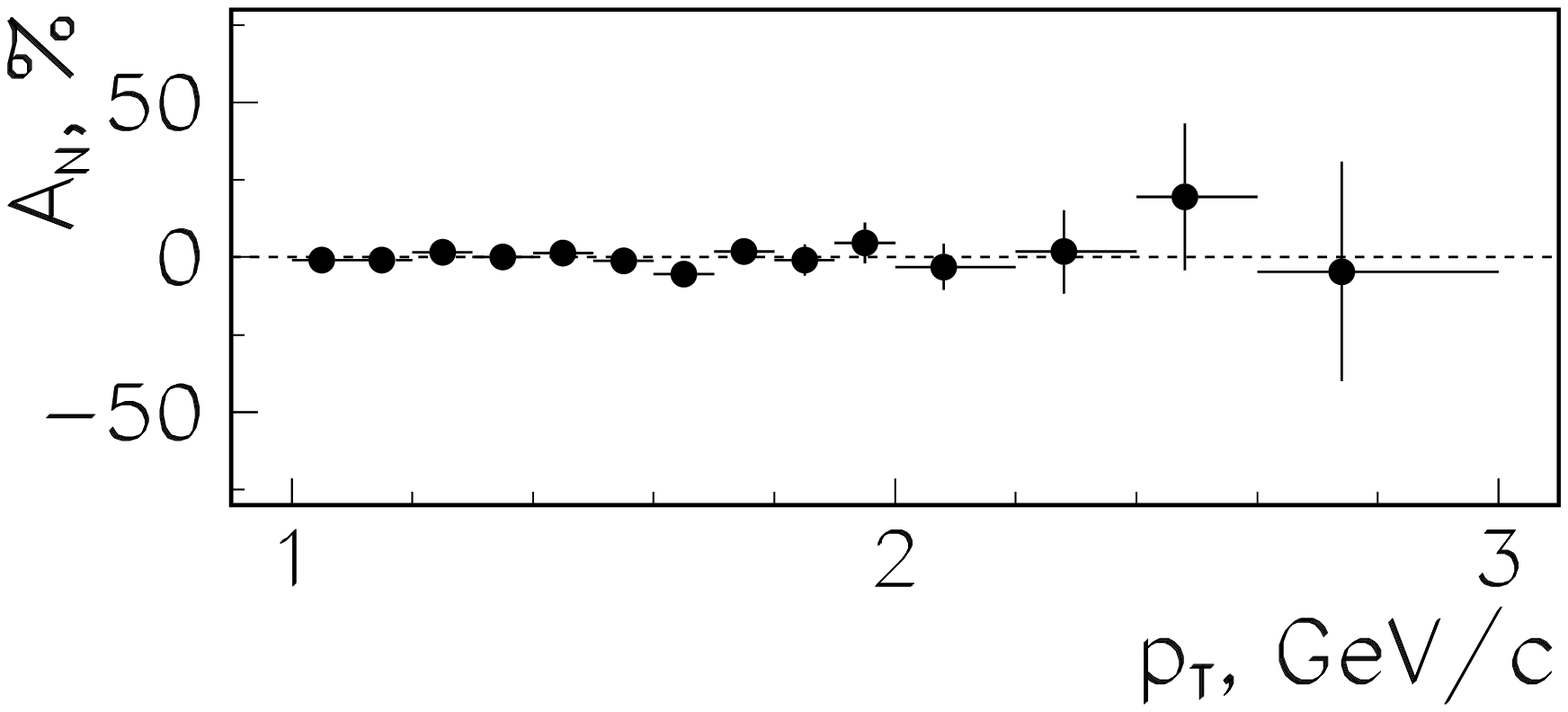,width=2.2in}}
\vspace*{-4mm}
\caption{$A_N$ in the reaction $\ppdup$ in the central 
region at 70~GeV.}
\label{fig:asym_centr}
\end{figure}

A large asymmetry was observed 
previously in the reaction $\pimp$ in the central region at 
40~GeV\cite{protvplb}. $A_N$ achieved -40\% at $p_T>2.5$~GeV/c. 
At the same time, the E704 experiment found zero asymmetry in 
the same kinematic region in the reaction $\pdupp$. 
Our recent measurements\cite{protv70cent} in the reaction 
$\ppdup$ at 70 GeV ({\bf Fig.~\ref{fig:asym_centr}}) are in 
agreement with E704 result. One may conclude that the 
asymmetry in the central region depends on the flavor of 
interacting particles and does not depend on the energy.

\begin{figure}[h]
\vspace*{-6mm}
\centerline{\psfig{file=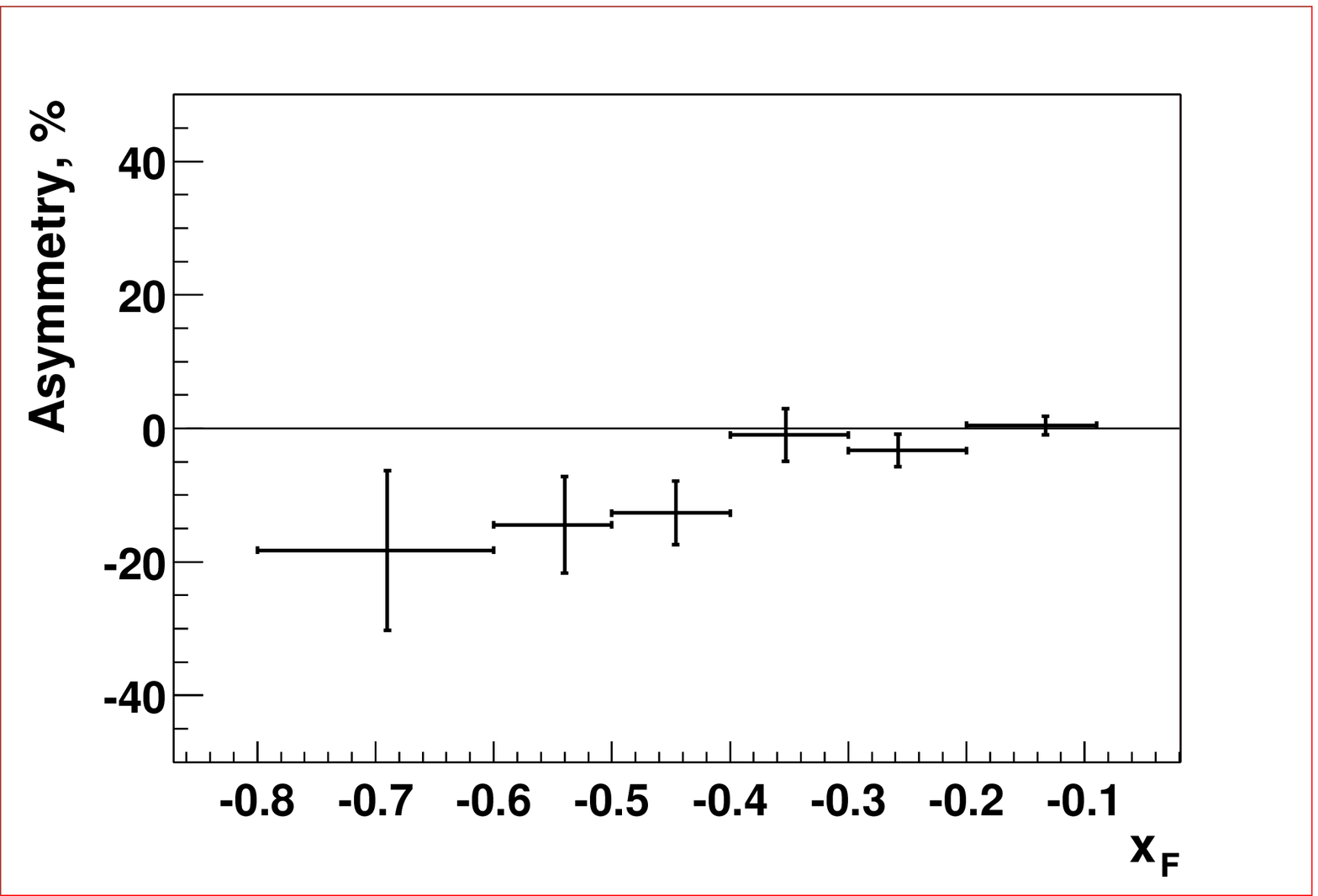,width=2.2in,height=1.2in}}
\vspace*{-2mm}
\caption{$A_N$ in the reaction $\pimp$ in the target 
fragmentation region at 40~GeV.}
\label{fig:asym_pi}
\vspace*{-4mm}
\end{figure}

Significant effects were observed eearlier in the polarized beam 
fragmentation region in the reaction $\pdupp$ at FNAL and BNL. 
We measured the asymmetry in the polarized target fragmentation 
region. $A_N$ in the reaction $\pimp$ 
at 40~GeV achieves $(-14 \pm 4)$\% (See {\bf Fig.~\ref{fig:asym_pi}}) 
at $-0.8 < \xf <-0.4$\cite{protv40}. 

\begin{figure}[h]
\vspace*{-6mm}
\centerline{\psfig{file=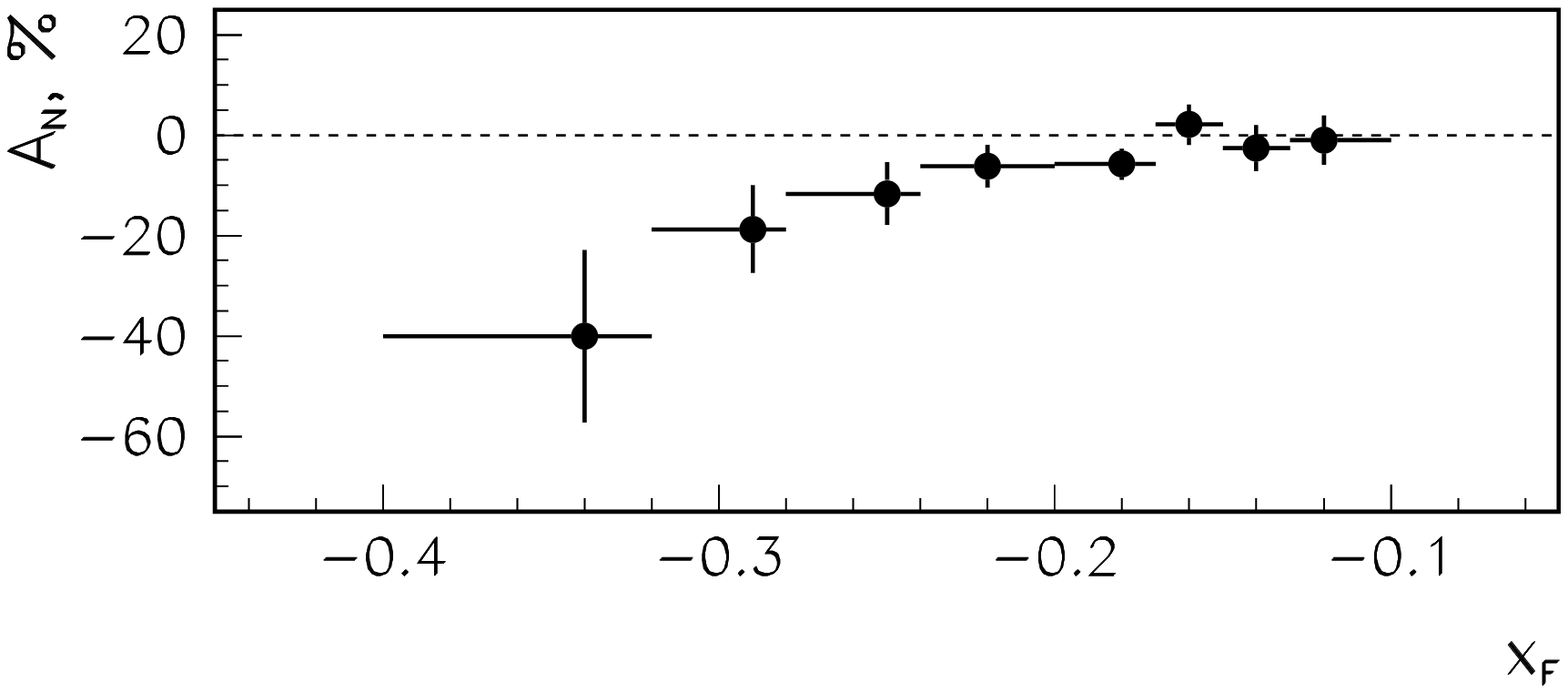,width=2.2in}}
\vspace*{-4mm}
\caption{$A_N$ in the reaction $\ppdup$ in the target 
fragmentation region at 70~GeV.}
\label{fig:asym_pback}
\vspace*{-8mm}
\end{figure}

The asymmetry in the reaction $\ppdup$ 
at 70~GeV equals to $(-24 \pm 8)$\% (See 
{\bf Fig.~\ref{fig:asym_pback}}) 
at $-0.4 < \xf <-0.28$\cite{protv70back}. The asymmetry in the 
polarized target fragmentation region does not depend on the
beam particle flavor.

\section{Discussion} 
\begin{figure}[t]
\centerline{\psfig{file=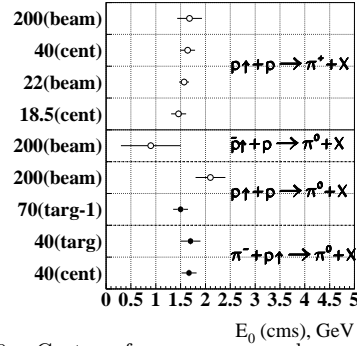,width=2.in}}
\vspace*{-4mm}
\caption{Center of mass energy where asymmetry 
starts to grow. The energy along the Y-axis is in GeV; 
$cent$ -- corresponds to experiments in the central region, 
$targ$ ($beam$) -- the polarized target (beam) fragmentation 
region.
}
\label{fig:scaling}
\vspace*{-2mm}
\end{figure}

The asymmetries in the both reactions are in an 
agreement with each other. $A_N$ in the polarized 
particle fragmentation region does 
not depend on the energy and on the particle flavor.
The result is in agreement with "universal threshold" of 
the asymmetry\cite{threshold} in the fixed target experiments 
(See {\bf Fig.~\ref{fig:scaling}}) and may be 
explained with the help of the constituent quark 
model\cite{const_quark}. 

\section*{Acknowledgments}
The work is partially supported by RFBR grant 06-02-16119.



\end{document}